\documentclass[aps,prb,showpacs,floatfix,amsmath,amssymb,superscriptaddress,reprint]{revtex4-2}
\usepackage{amsmath,amssymb} 
\usepackage{graphicx}
\usepackage{caption}
\usepackage{dcolumn}
\usepackage{color}
\usepackage{bm}
\usepackage{xurl}
\usepackage{url}
\usepackage{float}
\usepackage{balance}
\usepackage{adjustbox}
\usepackage[hypcap=false]{caption}  
\usepackage{stfloats} 
\usepackage[colorlinks=true,linkcolor=magenta,citecolor=magenta, urlcolor=blue]{hyperref}

\begin{document}

\title{Accelerating the Search for Superconductors Using Machine Learning}


\author{Suhas Adiga}
\email{suhasadiga@jncasr.ac.in}
\affiliation{Theoretical Sciences Unit, School of Advanced Materials (SAMat), Jawaharlal Nehru Centre for Advanced Scientific Research, Jakkur, Bengaluru 560064, India}
\affiliation{Chemistry and Physics of Materials Unit, \\ Jawaharlal Nehru Centre for Advanced Scientific Research, Jakkur, Bengaluru 560064, India}

\author{Umesh V. Waghmare}
\email{waghmare@jncasr.ac.in}
\affiliation{Theoretical Sciences Unit, School of Advanced Materials (SAMat), Jawaharlal Nehru Centre for Advanced Scientific Research, Jakkur, Bengaluru 560064, India}

\date{\today}

\begin{abstract}
Prediction of critical temperature (\(T_c\)) of a superconductor remains a significant challenge in condensed matter physics. While the BCS theory explains superconductivity in conventional superconductors, there is no framework to predict critical temperature of unconventional, higher $T_{c}$ superconductors. Quantum Structure Diagrams (QSD) were successful in establishing structure-property relationship for superconductors, quasicrystals, and ferroelectric materials starting from chemical composition. Building on the QSD ideas, we first demonstrate that the principal component analysis of superconductivity data naturally uncovers the clustering of various classes of superconductors. Earlier machine learning studies generally relied on datasets with inconsistencies and incomplete information, leading to a suboptimal predictions of superconductivity. To address this, we introduce a data-cleaning workflow to enhance the statistical quality of superconducting material databases by eliminating redundancies and resolving inconsistencies. With this improvised database, we apply a supervised machine learning framework and develop a Random forest model to predict superconductivity and critical temperature as a function of descriptors motivated from QSD. Our ML model demonstrates robust performance, achieving an \text{$R^{2}$} score of 0.87 and a mean absolute error of 5.6 K. We demonstrate that this model generalizes effectively in reasonably accurate prediction of $T_{c}$ of compounds outside the database. We further employ our model to systematically screen materials across multiple materials databases as well as various chemically plausible combinations of elements and predict $\mathrm{Tl}_{5}\mathrm{Ba}_{6}\mathrm{Ca}_{6}\mathrm{Cu}_{9}\mathrm{O}_{29}$ to be a superconductor with a $T_{c}$ $\sim$ 105 K. Being based on the descriptors used in QSD's, our model bypasses structural information and predicts $T_{c}$ merely from the chemical composition, and hence is quite efficient and powerful in the search for new superconductors.
\begin{description}
\item[Keywords]
\textit{Superconductivity, Machine learning, Critical temperature} 
\end{description}
\end{abstract}

\maketitle
\vspace{-0.25cm}
\section{\label{sec:level1} Introduction}
\vspace{-0.25cm}
The discovery of superconductivity in mercury (Hg) was serendipitous, occuring below a critical temperature of 4.2 K and first observed by Onnes in 1911 \cite{Onnes1911}. Superconductors are materials that conduct electric current with zero resistance below the critical temperature (\( T_{\textnormal{c}} \)). They are used in number of applications in Magnetic Resonance Imaging (MRI), Maglev trains, particle accelerators, energy transmission and as qubits for next-generation quantum computers. Shortly after the discovery of superconductivity in elements like Hg and Pb, it was also observed in alloys \cite{Smith1935237}, A15 compounds \cite{Hardy1953}, cuprates \cite{Bednorz1986}, iron-based pnictides \cite{Kamihara2006}, and metal hydrides \cite{Satterthwaite1970}. Currently, the highest $T_c$ achieved at ambient pressure is 138 K in Hg-based cuprates \cite{Schilling1993}, which increases to 160 K under a pressure of 23 GPa \cite{Gao1994}. Ashcroft argued that metallic hydrogen, under high pressure, could exhibit even higher critical temperature \cite{Ashcroft1968}. To date, the highest \( T_{\textnormal{c}} \) is observed in La$\textnormal{H}_{10}$, which shows near-room-temperature superconductivity at 260 K, albeit under an extremely high pressure of 190 GPa \cite{Somayazulu2019}.

Superconductors are broadly of two types: Conventional and Unconventional superconductors. Conventional superconductors are those where the mechanism of superconductivity is based on electron-phonon coupling within the BCS Theory \cite{Bardeen1957}, where phonons provide attractive interactions between two electrons that form a Cooper pair. Compounds like intercalated graphite (e.g. $\text{C}_{6}\text{Yb}$  \cite{Weller2005} ), A15 compounds (e.g. $\text{V}_{3}\text{Si}$  \cite{Hardy1953} ), elemental compounds (e.g. Hg \cite{Onnes1911} ), and metal hydrides (e.g $\text{CeH}_{9}$ \cite{Li2019} ) belong to this class. Eliashberg Theory, an extension of the BCS theory that incorporates the full electron-phonon interaction \cite{Eliashberg1960}, complemented by the empirical equation proposed by McMillan \cite{McMillan1968} and Allen-Dynes  \cite{Allen1975} estimates the critical temperature of superconductors based on electron-phonon coupling and average phonon frequency. 

This formalism can be readily used within the framework of first-principles Density Functional Theory to predict critical temperature \cite{COHEN2010} of conventional superconductors. Among the known conventional superconductors at ambient pressure, the highest critical temperature is observed in Magnesium Diboride ($\textnormal{MgB}_{2}$) with a critical temperature of 39 K \cite{Nagamatsu2001}. 

High temperature superconductors like cuprates, iron-based compounds are unconventional superconductors, where there are various mechanisms like hole theory \cite{Hirsch1989}, resonating valence bond theory \cite{Anderson1987}. However, these mechanisms do not quantitatively predict critical temperature. While achieving superconductivity at room temperature is a dream of condensed matter physicists, developing a predictive model of critical temperature remains a serious challenge. The high computational costs for screening thousands of compounds using Density Functional Theory (DFT) makes it resource intensive. Additionally, DFT based methods are limited to predict critical temperature in a weakly correlated conventional systems. This hinders the exploration for materials with potential for room temperature superconductivity. 

Statistical methods have been powerful tools, even before the conceptualization of machine learning. One such statistical approach was developed in the 1980's by J.C. Phillips. It involved using Quantum Structural Diagrams \cite{Phillips1991} to establish structure-property relationships based on chemical composition. 

Many attempts have been made since 2018 to estimate critical temperature using machine learning. Stanev et al.\cite{Stanev2018} used 145 Magpie descriptors and 26 Aflow features based on elemental composition to predict critical temperature using Random forest method. Hamidieh's \cite{Hamidieh2018} work presented a data-driven statistical model to predict critical temperature of superconductors using only the chemical formulas of the materials. This model extracted 81 features based on the elemental properties. 

Gashmard et al.\cite{Gashmard2024} used 322 atomic features generated from Soraya package in Python to predict critical temperature of superconductors. Zhang et al.\cite{Zhang2022} used machine learning with structural descriptors and electronic properties from the Aflow database to predict the critical temperature of superconductors. Jung et al.\cite{Jung2024} developed a framework that uses gradient-boosting and statistical feature selection to predict critical temperature. Trezza et al.~\cite{Trezza2023} developed a machine learning classification framework to distinguish superconductors above and below temperature thresholds of 15 K (for benchmarking) and 35 K (for mixed-feature optimization). Their model was based on the Matminer descriptors~\cite{Ward2018} and incorporated several techniques, including invariant group identification and SHAP for feature reduction. Classification was performed using entropy-based Quasi Equilibrium Grids and Extra Trees. In their later work \cite{Trezza2024}, they extended this approach to address sampling bias between specialized databases like SuperCon \cite{SuperCon2011} and general-purpose databases like the Materials Project (excluding SuperCon entries)~\cite{Jain2013}, thereby enabling validation of $T_c$ predictions, even though the latter may include materials not tested for superconductivity. 

Konno et al.\cite{Konno2021} introduced a novel deep learning model that learns to interpret the periodic table, rather than relying on manually engineered features. This model was trained on the materials from SuperCon and Crystallographic Open Database (COD). Kaplan et al.\cite{Kaplan2025} employed a convolutional neural network (CNN) to analyze the chemical compositions of superconductors without augmenting non-superconductors in their model. Pereti et al.\cite{Pereti2023} utilized a Deep Set neural network that processes chemical compositions as sets, thus eliminating the bias from the ordering of the elements. The model was trained on data from the SuperCon and Crystallographic Open Database (COD) databases.

Choudhary et al.\cite{Choudhary2022} presented the j-Scr workflow, which combines DFT-based screening with empirical rules and deep learning to predict superconductor transition temperatures. The workflow relies on the McMillan-Allen-Dynes formula. Cerqueira et al.\cite{Cerqueira2023} combined \textit{ab-initio} calculations of electron-phonon coupling using Density-Functional Perturbation Theory (DFPT) with machine learning to predict critical temperature in conventional superconductors. The use of the machine learning model MODnet allowed them to reduce the number of calculations by focusing on the promising candidates in the material search. Seegmiller et al.\cite{Seegmiller2023} introduced the DiSCoVeR algorithm and emphasized the importance of a composition-based approach for identifying new superconductors.  

In this work, we first focus on developing a cleaner dataset of superconducting materials. Secondly, we use chemical composition-based descriptors derived from the Quantum Structure Diagrams~\cite{Phillips1991}. We finally use classical machine learning algorithms to predict superconductivity and critical temperature.

In Sec \ref{sec:Feature_Engineering}, we discuss on features and descriptor generation from chemical composition using Quantum Structure Diagrams \cite{Phillips1991} for training a machine learning model. In Sec \ref{sec:Methodology}, we describe our data cleaning workflow for preparing the \textit{SuperCon-MTG} dataset. Section~\ref{sec:Results} presents an analysis of the data distribution in our modified database, a Principal Component Analysis (PCA) study of superconductors, and the use of a Random forest model for classification and regression tasks. We present SHAP \cite{SHAP} analysis to interpret the role of each feature in predicting the critical temperature, identify the key features and finally train a model using only these features. In Sec.~\ref{sec: New Materials}, we demonstrate the search for new superconducting materials and validate our model's performance with recently discovered compounds and exploring potential superconductors in the Materials Project \cite{Jain2013} database.

\vspace{-0.25cm}
\section{\label{sec:Feature_Engineering} Descriptors from QSD's and Engineered features}
\vspace{-0.25cm}
Machine learning models rely on features that effectively predict the target variable. Superconducting critical temperature can depend on various factors. In conventional superconductors, factors like electron-phonon coupling constant ($\lambda$), Debye temperature ($\theta_{\textnormal{D}}$) play important roles. In unconventional superconductors factors such as hole doping, Cu-O bond lengths become significant. 

Before the advent of BCS theory in the 1950s, the discovery of superconductors relied on empirical rules. One such rule was given by Matthias \cite{Matthias1957, Matthias1970}, who identified over a hundred superconducting materials, primarily alloys, by following heuristic principles. Among these, the average valence electron count per atom (Z) emerged as an important predictor. Superconductivity was observed only when Z was between 2 and 8. This rule suggested that materials with high symmetry, preferably cubic or hexagonal, are plausible candidates for superconductivity. However, Matthias' rule failed to account for the absence of superconductivity in metals like Mg, which has Z = 2, and in La-Ba-Cu-O materials which possess a layered structure. 

The three golden descriptors of Quantum Structural diagrams are: (a) Weighted average of metallic electronegativity difference ($\overline{\Delta \chi}$), (b) Weighted average of orbital radii difference ($\overline{\Delta R}$) and (c) Weighted average of valence electron number ($\overline{N_v}$). It was successful in organizing crystal structure types of binary and ternary alloys \cite{Villars1987}, as well in identifying new quasicrystals \cite{Villars1986}. Villars et al. \cite{Villars1988} showed that for 60 known superconductors with \( T_{\textnormal{c}} \) $> 10 \text{K}$, these superconductors were found in one of the three islands in \( T_{\textnormal{c}} \) v/s $\overline{N_v}$ plane. The first island was dominated by the A15 compounds, the second by the borides, carbides and nitrides and the third one by the Cu-oxide, chevrel sulphide and selenide superconductors. Later, Rabe et al. \cite{Rabe1992} extended this analysis to 70 known superconductors at that time and uncovered a similar pattern. Using Quantum Structural Diagrams, they uncovered that oxide ferroelectric materials were localized in a smaller region. This was significant because it was not merely a consequence of specific structural types but rather a reflection of the underlying chemical factors responsible for ferroelectricity. Furthermore, Balasubramanian et al.~\cite{Bala1989} provided a physical rationale for using electronegativity as a meaningful descriptor. They argue that superconductivity results from the unimpeded motion of Cooper pairs, which requires a near-zero net field, a condition that can hypothetically occur in flat lattice potentials. Since electronegativity governs how atoms attract electrons, it can serve as a proxy. Inspired by these, we explore if these descriptors can be used within machine learning models to predict critical temperature of superconductors.

In the Quantum Structure Diagram schemes, descriptors of a binary compound ($\text{A}_{m}\text{B}_{n}$) with a normalized composition ($\text{A}_{x}\text{B}_{y}$) where  $x = \frac{m}{m+n}$, $y = \frac{n}{m+n}$ , $x \leq y$ and $x + y = 1$ are defined as: 
\begin{equation}
    \overline{\Delta {\chi}} = 2x(\chi_A - \chi_B)
\end{equation}
\begin{equation}
    \overline{\Delta{R}} = 2x(R_A - R_B)
\end{equation}
\begin{equation}
     \overline{N_v} = xN_A + yN_B
\end{equation}
where `$\chi$' is the electronegativity in Martynov - Batsanov scale, `$R$' is the Zunger's pseudopotential radii sums, `$N$' is the valence electron number (see Supplementary Information (SI), Sec. \textcolor{magenta}{I} \cite{SI}), $\overline{\Delta \chi}$ is the weighted average of metallic electronegativity difference, $\overline{\Delta R}$ is the weighted average of orbital radii difference, $\overline{N_v}$ is the weighted average of valence electron number and subscripts denotes the species `A' and `B'.       

For a ternary compound \(\text{A}_{m}\text{B}_{n}\text{C}_{t}\), with a normalized composition \(\text{A}_{x}\text{B}_{y}\text{C}_{z}\), where \(x \leq y \leq z\), x + y + z = 1 and 
\[
\quad x = \frac{m}{m+n+t}, \quad y = \frac{n}{m+n+t}, \quad z = \frac{t}{m+n+t},
\]
the Quantum Structure Diagrams-based descriptors are given as:
\begin{equation}
    \overline{\Delta {\chi}} = 2x(\chi_A - \chi_B) + 2x(\chi_A - \chi_C) +  2y(\chi_B - \chi_C)
\end{equation}
\begin{equation}
    \overline{\Delta{R}} = 2x(R_A - R_B) + 2x(R_A - R_C) + 2y(R_B - R_C)
\end{equation}
\begin{equation}
     \overline{N_v} = xN_A + yN_B + zN_c
\end{equation}

Quantum Structure Diagrams require the analysis of quaternary and higher-order compounds, treating them as pseudo-ternary compounds by grouping similar elements. However, we extend this approach to treat quaternary and higher-order compounds in a manner similar to binary and ternary compounds. When calculating the weighted average of the property difference, each possible pair is weighted by twice the minimum of their normalized compositions. For the weighted average of a property, the normalized composition of each element is used as a weight.

Building on this intuition, we introduce additional descriptors, including:  
\begin{itemize}  
    \item The weighted average of the unpaired electron number (\(\overline{U_e}\)),  
    \item The weighted average of electronegativity (\(\overline{\chi}\)),  
    \item The weighted average of orbital radius (\(\overline{R}\)), and  
    \item 24 statistical features using properties weighted by the normalized composition, including the median, maximum, minimum, range, standard deviation, and average deviation of electronegativity, orbital radius, valence electron number, and unpaired electron number.   
\end{itemize}  

For a binary compound with normalized composition ($\text{A}_{x}\text{B}_{y}$). The descriptors are calculated as follows:
\begin{equation}
    \overline{U_e} = xU_A + yU_B
\end{equation}
\begin{equation}
    \overline{\chi} = x\chi_{A} + y\chi_{B}
\end{equation}
\begin{equation}
    \overline{R} = xR_A + yR_B
\end{equation}

To determine the statistical features of a property, we create a one-dimensional array ($W_{\text{comp}}$) by weighting over the normalized composition and evaluate each statistical feature using this array. For a ternary compound ($A_{x}B_{y}C_{z}$), the $W_{\text{comp}}$ array for electronegativity is:
\begin{equation}
    W_{\text{comp}} = \left[ x \chi_A, y \chi_B, z \chi_C \right]
\end{equation}

Thus, our feature space consists of six descriptors and 24 statistical features related to four different elemental properties. While superconductivity depends on various factors, including preparation methods, applied pressure and magnetic fields. Our database reports the critical temperature at ambient pressure. As these descriptors do not capture pressure dependence of critical temperature, our model targets prediction of critical temperature at ambient conditions. 

Our goal is to develop a model for prediction of a new superconductor. As prediction of structure from chemical composition itself is a daunting problem, we use only composition based descriptors here in training a machine learning model to predict critical temperature.

\vspace{-0.25cm}
\section{\label{sec:Methodology} Methodology}
\vspace{-0.25cm}
\subsection{\label{subsec: Data Cleaning} Preprocessing and Data Cleaning}
\vspace{-0.25cm}
A \textit{clean} database is essential to analysis with any machine learning algorithm \cite{Pankajakshan2017}. The notion of ``Garbage in, garbage out" holds true, if a model is trained on flawed data, its predictions are bound to be unreliable. In materials science research, data availability is often a significant challenge, constrained by small databases. 

Here, we use the SuperCon database maintained by NIMS, Japan \cite{SuperCon2011}, which was initiated following the discovery of high-\(T_c\) superconductors. Three versions of this database are currently available, primarily containing information on the chemical formula, \(T_c\), the reporting journal and structural details. The database also includes measurement methods such as \(T_c\) onset, the temperature at which the superconducting transition begins, \(T_c\) at zero resistance, the temperature at which zero resistance is achieved, and magnetization studies for some compounds. The SuperCon database cleaned and published by Stanev et al. \cite{Stanev2018} in 2018 includes 16,414 compounds. A later version, \textit{SuperCon - Ver.220808} \cite{SuperCon220808}, contains 33,406 compounds, while the most recent release, \textit{SuperCon - Ver.240322} \cite{SuperCon240322}, includes 26,358 compounds. Both the 2018 \cite{Stanev2018} and 2024 \cite{SuperCon240322} databases, have no missing \(T_c\)'s. Unlike, \textit{SuperCon - Ver.220808}\cite{SuperCon220808}, which contains missing entries for \(T_c\).
\onecolumngrid
\begin{center}
    \includegraphics[width=0.66\linewidth]{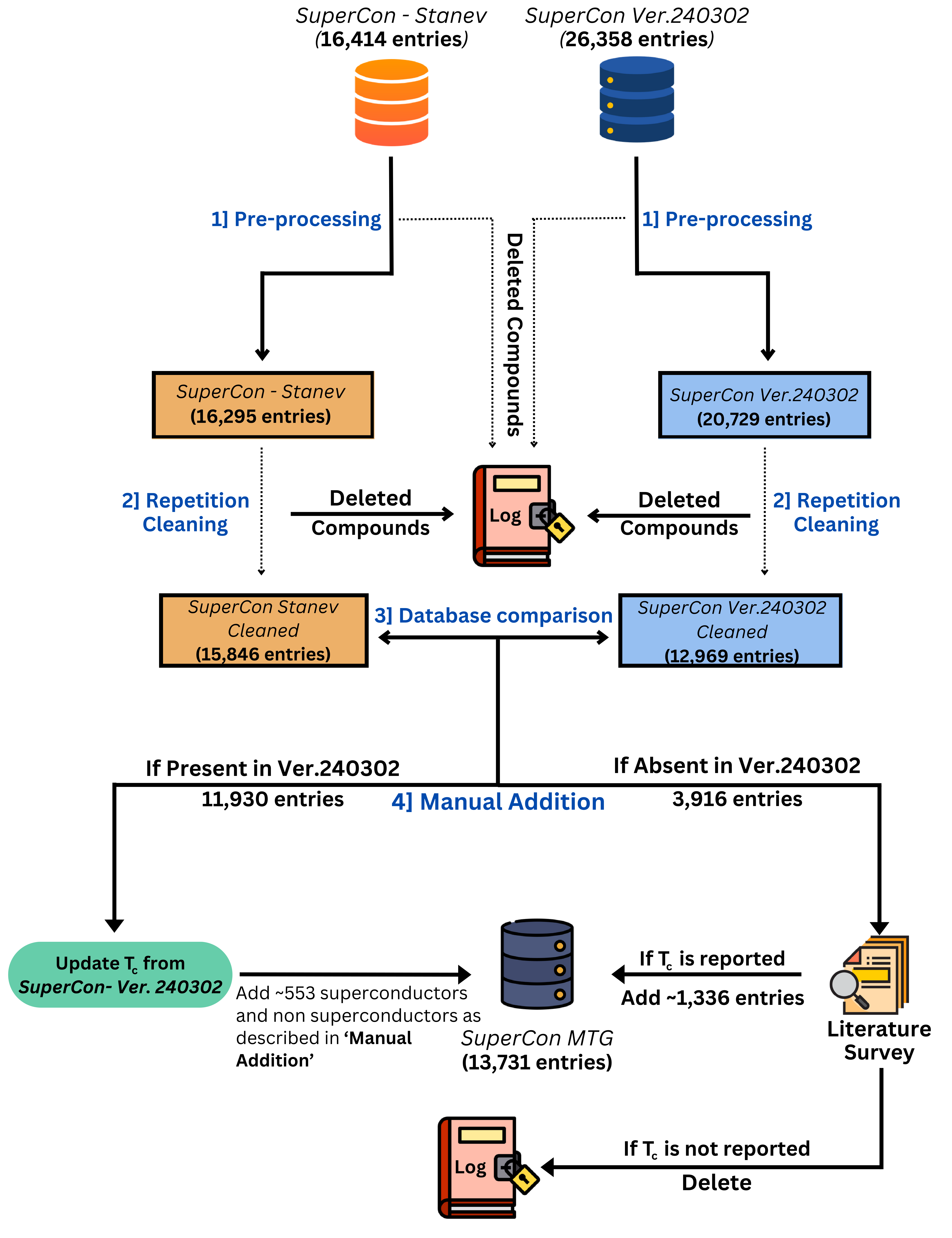}
    \captionof{figure}{A schematics of data cleaning workflow employed in this study for preparation of \textit{SuperCon-MTG}.}
    \label{fig:Data_Cleaning}
\end{center}
\twocolumngrid
The data-cleaning algorithm consists of four steps: pre-processing, repetition cleaning, database comparison and manual addition of compounds to the database. Fig.~\ref{fig:Data_Cleaning} shows graphical representation of our data cleaning workflow.
\begin{enumerate}
    \item \textbf{Pre-processing} : This step ensures that all compounds in the database follow the required format of chemical formula to facilitate feature generation. Since our model relies solely on composition-based features, we focus on cleaning the database to ensure the validity of chemical compositions.
    \begin{itemize}
        \item If an element appears more than once in a chemical formula, we consolidate its composition at the first occurrence of that element.
        \\
        E.g., $\text{Nb}_3\text{Sn}_{0.2}\text{Sn}_{0.8}\xrightarrow{} \text{Nb}_{3}\text{Sn} $ 
\item Compounds with incorrect chemical formulas were identified manually. Their formulas were corrected by referring to the cited journal. 
\textit{E.g.}, $\text{La}_{0.75}\text{Pr}_{0.25}\text{Ba}_{22}\text{O}_{8} \rightarrow \text{La}_{0.75}\text{Pr}_{0.25}\text{Ba}_{2}\text{O}_{8}$ and $\text{Os}_{10} \rightarrow \text{Os}$.

\item We remove compounds with variable or invalid stoichiometric numbers, as well as those containing unsupported characters such as `+', `-', `!', or `=', since these cannot be practically used for feature generation. 
\\
E.g., $\text{EuBa}_{2}\text{Cu}_{3}\text{O}_{Y}$, $\text{Sr}_{2}\text{Ca}_{4}\text{Cu}_{5}\text{O}_{2}\text{N}_{+3}\text{B}$, $\text{La}_{2}\text{Sr}_{0}\text{Cu}_{0.95}\text{Zn}_{0.05}\text{O}_{4}$.
        \item Compounds with a stoichiometric number of 150 or greater were removed, as such unbalanced systems are rarely found.
        \\
        E.g., $\text{Y}\text{Ba}_{2}\text{Cu}_{3}\text{O}_{6050}$
    \end{itemize}
    This process is applied consistently to all the entries in the databases.
    \item \textbf{Repetition Cleaning}: Since our feature generation process involves normalization of the composition, different representations of the same compound become indistinguishable. For example $\text{AB}_{3}$ and $\text{A}_{0.25}\text{B}_{0.75}$ are indistinguishable.  We identify such entries in the database and average out their critical temperature values.  For \textit{SuperCon-Stanev} database we simply take the averaged critical temperature. For \textit{SuperCon - Ver.220808} and \textit{SuperCon - Ver.240322} we retain the compound with averaged \( T_{\textnormal{c}} \) only if their standard deviation is less than 5 K (see Supplementary Information Sec. \textcolor{magenta}{II} \cite{SI} for details). In the \textit{SuperCon-Stanev} database, we do not apply this standard deviation filter, as critical temperatures in this database will later be updated using the most recent data from \textit{SuperCon - Ver.240322}.

    \item \textbf{Database Comparison}: Comparison of the cleaned versions of \textit{SuperCon - Ver.220808} and \textit{SuperCon - Ver.240322}, shows that all 12,950 compounds in \textit{SuperCon - Ver.220808} are present in \textit{SuperCon - Ver.240322}. Similarly, on comparing  \textit{SuperCon-Stanev} with that of \textit{SuperCon - Ver.240322}, 11,930 compounds in \textit{SuperCon-Stanev} have critical temperature reported in both the databases. The remaining 3,916 compounds have reported critical temperature of 0 K, but are absent in the latest version of the database. Manual search of these compounds shows that some are superconductors, while others are not. It seems that the database of Stanev et al. \cite{Stanev2018} has imputed the missing data, which is not a good approach for ML model, which would learn from wrong data. Hence, we do manual search of such compounds and update their data back to the database as explained in the next step.
    
    \item \textbf{Manual Addition}: We manually checked the cited journals for the 3,916 compounds with reported critical temperature of 0 K. A compound is added back to the database only if the source explicitly confirms its superconducting nature or lack thereof.
    \begin{itemize}
        \item Many compounds exhibit superconductivity alongside magnetic ordering, charge density waves, or spin density waves. We retain a compound in the database only if it is observed to exhibit superconductivity or no superconductivity.
        \\
        E.g, $\textnormal{LaFeAs}_{0.3}\text{P}_{0.7}\text{O}$ \cite{Lai2014}, the phase diagram reported in the cited paper identifies it as an antiferromagnet without superconductivity, We thus update its \( T_{\textnormal{c}} \) as 0 K.  
        \item As absolute zero is unattainable, some research articles report that a compound might be superconducting below the lowest attainable temperature of the apparatus. If such lower limit is 0.1 K or lower, we approximate its \( T_{\textnormal{c}} \) as 0 K.
        \item Many published works do not report numerical value of \( T_{\textnormal{c}} \) for a given composition but provide magnetization, resistivity, specific heat studies, or phase diagrams plots. We manually analyzed over 150 images and interpolated critical temperature from these graphs. In cases where multiple studies report different \( T_{\textnormal{c}} \) values for the same compound, we recorded their average.
        \\
        By following this criterion, we were able to add 1,337 of the 3,916 compounds back into the database.
        
        \item A machine learning model will be biased if it is trained only on superconductors. To the cleaned database which contains about 1,100 non-superconductors, we add 455 more non-superconductors, along with 43 superconductors reported by Hosono et al \cite{Hosono2015}. Additionally, through our review of the literature, we identified and manually added 55 materials exhibiting both superconducting and non-superconducting behavior that were not present in the original database. Upon addition of these compounds, some repetition arose, we average the critical temperature of such repeated compounds and retain them only if the standard deviation is less than 5 K. This final cleaned database is referred to as `\textit{SuperCon-MTG'} named after the `\textit{Materials Theory Group}'. \textit{SuperCon-MTG} is a cleaned and augmented version of the \textit{SuperCon} database originally developed and maintained by NIMS Japan \cite{SuperCon2011, SuperCon240322}. The original database is available under the Creative Commons Attribution 4.0 International (CC BY 4.0) license.
       
    \end{itemize}
\end{enumerate}
\onecolumngrid
\begin{center}
    \includegraphics[width=0.89\linewidth]{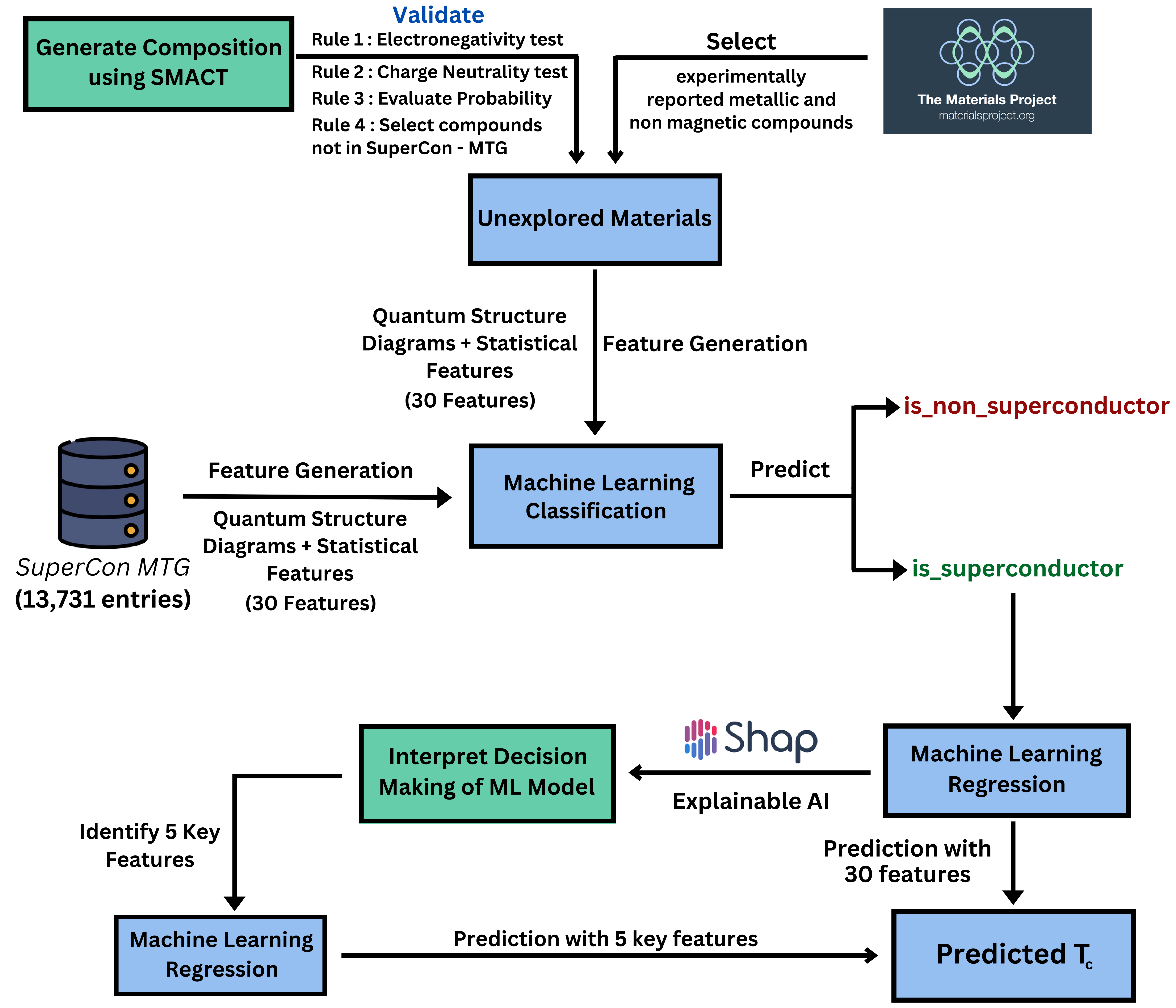}
    \captionof{figure}{A schematic of Machine learning workflow}
    \label{fig:ML_Workflow}
\end{center}
\twocolumngrid
\vspace{-0.25cm}
\subsection{\label{subsec: Machine Learning Workflow} Machine Learning Workflow}
\vspace{-0.25cm}
Non-superconductors are considered to have a critical temperature of 0 K in our database. However, accurately predicting 0 K using regression-based machine learning algorithms is challenging. To address this, we strategize our workflow by performing a classification step followed by regression on the predicted superconductors.  

The SuperCon-MTG dataset used for training the model is split in an 80–20 ratio for training and testing, respectively (see Supplementary Information Sec. \textcolor{magenta}{III} \cite{SI} for details). Specifically, we first classify whether a material is a superconductor or a non-superconductor. If a compound is predicted to be a superconductor, we apply regression-based machine learning models to estimate its critical temperature. Otherwise, we assign it a critical temperature of 0 K. This two-step approach enhances the reliability of our machine learning predictions. Fig.~\ref{fig:ML_Workflow} provides a graphical representation of our machine learning workflow.

\vspace{-0.25cm}
\section{\label{sec:Results} Results}
\vspace{-0.25cm}
After data cleaning with the algorithm described in Sec.~\ref{subsec: Data Cleaning}, the \textit{SuperCon-MTG} dataset hosts 13,731 compounds, including both reported superconductors and non-superconducting compounds. The cleaned database contains 1,584 non-superconducting compounds with a critical temperature of 0 K. The highest reported critical temperature in the dataset is 143 K for $\text{Hg}_{0.66}\text{Pb}_{0.34}\text{Ba}_{2}\text{Ca}_{1.98}\text{Cu}_{2.97}\text{O}_{8.4}$ \cite{SHAO1995}. The average critical temperature of the compounds in the dataset is 21.2 K. The first quartile (\textit{$Q_{1}$}) lies below 2.05 K, meaning that 25\% of compounds in the database have critical temperature $<$ 2.05 K. The second quartile (\textit{$Q_{2}$}) is below 7.5 K and third quartile (\textit{$Q_{3}$}) falls below 29.6 K. 

A clear class imbalance exists between superconductors and non-superconductors, as the number of superconducting materials is significantly higher. This imbalance arises due to limited availability of labeled non-superconducting materials data in materials science. Despite the existence of well-maintained theoretical and experimental databases such as \textit{Materials Project} \cite{Jain2013} which hosts over 150,000 inorganic materials, \textit{The Open Quantum Materials Database (OQMD)} \cite{Saal2013}, which contains over 1.2 million DFT calculated thermodynamic and structural properties and the \textit{Inorganic Crystal Structure Database (ICSD)} \cite{Belsky2002} hosting more than 300,000 experimetally determined compounds, none of these databases explicitly label materials as superconducting / non-superconducting. Due to this limitation, we cannot manually add materials from these databases to balance the classes, which makes it difficult to ensure a well-distributed dataset for machine learning models.
\onecolumngrid
\begin{center}
    \includegraphics[width=0.71\linewidth]{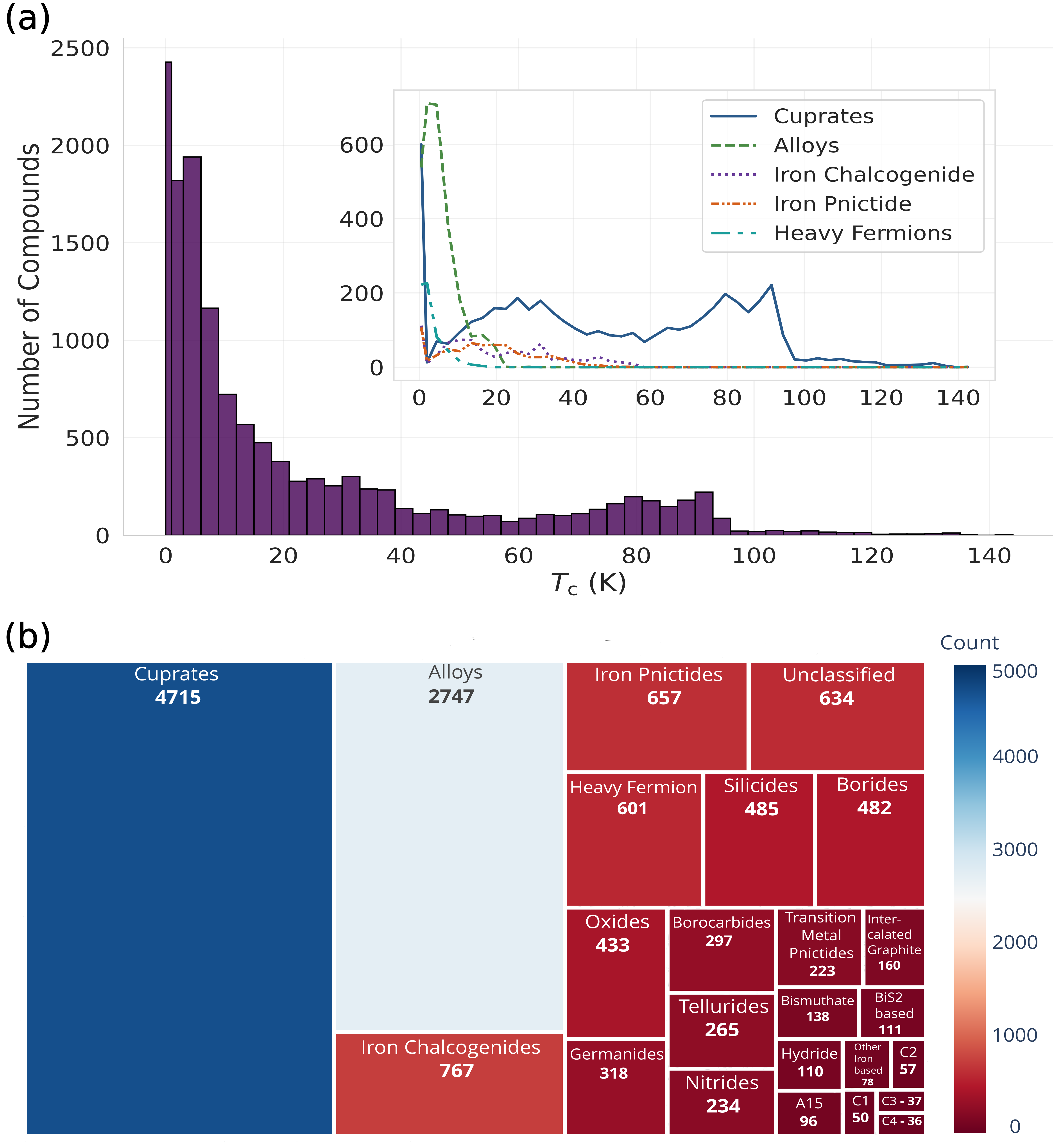}
    \captionof{figure}{(a) Histogram of \(T_c\) in \textit{SuperCon-MTG}, where the first bin (at \(T = 0\) K) has a width of 1 K, while all other bins have a width of 3 K with inset of distribution of \(T_c\) for 5 largest classes of superconductors. (b) The distribution of compounds in the SuperCon-MTG dataset by class, including Cuprates, Alloys, Iron Chalcogenides, Iron Pnictides, Heavy-Fermion materials, Silicides, Borides, Oxides, Borocarbides, Tellurides, Germanides, Nitrides, Transition Metal Pnictides, Intercalated Graphite, Hydrides, Bismuthates, Elemental Superconductors (labeled as C2), A15 Compounds, Fullerides (labeled as C1), $\text{BiS}_2$-based materials, other Iron-based Compounds, Chevrel phases (labeled as C3), Dichalcogenides (labeled as C4), and Unclassified materials.}
    \label{fig:Hist_Tc}
\end{center}
\twocolumngrid
We classify the compounds in \textit{SuperCon-MTG} into twenty-four classes based on their chemical composition. Cuprates (copper oxide compounds) constitute the largest fraction, accounting for 34.3\% of the database, followed by alloys at 20\%. Iron chalcogenides make up approximately 5.6\%, Iron pnictides 4.8\%, and Heavy Fermion compounds around 4.4\%. Unidentified class of compounds fall under `Unclassified' constituting 4.6\%. The remaining 18 classes collectively contribute 26.3\% to the dataset. The extensive research on cuprates, driven by their potential as high $T_c$ superconductors, is evident from their dominant presence, making up nearly one-third of the dataset. 

\vspace{-0.25cm}
\subsection{\label{subsec : PCA} Unsupervised Learning: Principal Component Analysis}
\vspace{-0.25cm}
Previous studies have demonstrated the clustering of compounds in the \( T_c \) vs. \(\overline{N_v} \) space \cite{Villars1988, Rabe1992}. This clustering arises because certain classes of materials exhibit superconducting critical temperatures within specific ranges and tend to share common crystal structures. As a result, clustering in this space is a natural consequence rather than an insightful distinction. In our analysis, we extend this idea by projecting superconducting materials into two principal components using Principal Component Analysis (PCA). Two of the three QSD-based descriptors, weighted average of valence electron number and weighted average of orbital radius difference are the 9th and 10th most important features influencing the PC-1 component. We observe distinct clustering patterns, with compounds grouping according to their structural and chemical similarities. Notably, cuprates, elemental superconductors, and carbon-based compounds exhibit clear clustering. Other classes also cluster, though with some degree of overlap between different groups, making their boundary diffused. Clustering of these compounds in PCA space can be seen as in Fig.~\ref{fig:PCA}.
\onecolumngrid
\begin{center}
    \includegraphics[width=1\linewidth]{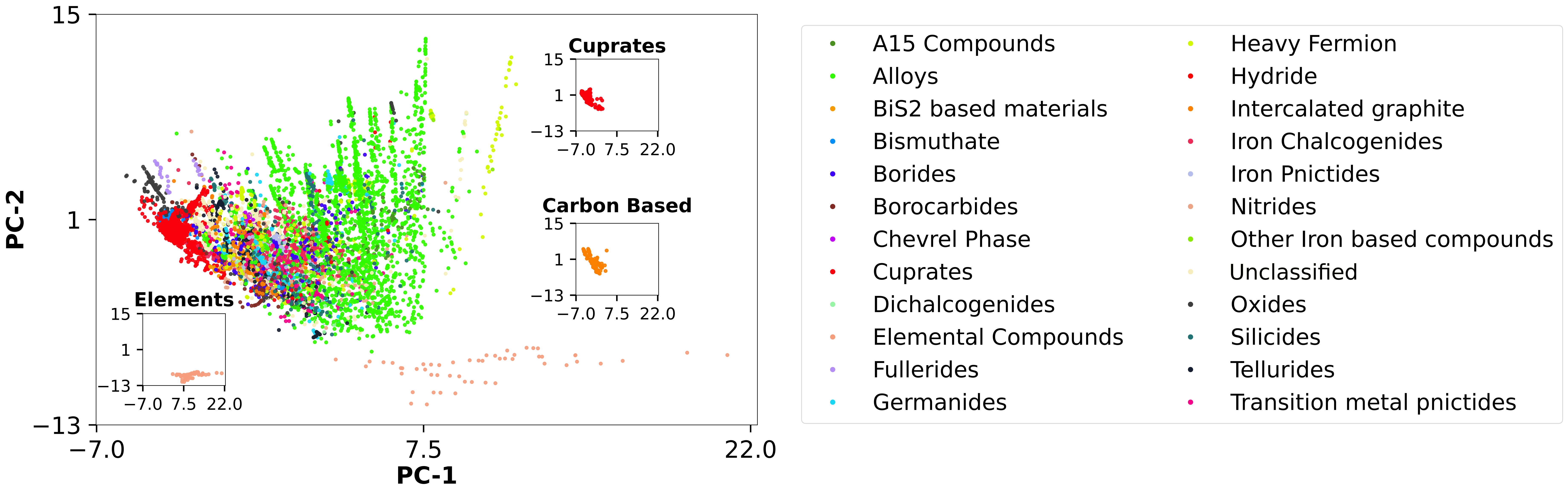}
    \captionof{figure}{Clustering of superconducting materials in first two principal components (PC-1 and PC-2) space using the Principal Component Analysis (PCA). Distinct clusters corresponding to cuprates, elemental superconductors, and carbon-based compounds are evident and are shown as an inset in the plot.}
    \label{fig:PCA}
\end{center}
\twocolumngrid

\vspace{-0.25cm}
\subsection{\label{subsec: Hyperparameter Tuning} Hyperparameter Tuning}
\vspace{-0.25cm}
Random forest \cite{Breiman2001} is an ensemble learning method that constructs multiple decision trees using bagging. Specifically, it selects random samples with replacement (bootstrapping) from the dataset to train each tree. This means that after a sample is drawn from the dataset, it is returned to the bag, making it possible for the same sample to be selected multiple times for training different trees. For regression tasks, the final output is the average prediction of individual trees, whereas for classification, it is determined by majority voting. This random sampling process helps reduce overfitting and variance.

The Random forest model consistently outperforms both XGBoost \cite{Chen2016} and Extra Trees for classification and regression tasks related to superconductivity prediction (see Supplementary Information Sec. \textcolor{magenta}{III} \cite{SI} for details). The overall workflow is outlined in Sec \ref{subsec: Machine Learning Workflow}.

\vspace{-0.25cm}
\subsubsection{\label{subsubsec: Classification}Classification}
\vspace{-0.25cm}
We perform hyperparameter tuning to select the best parameters for fitting our Random forest classification model using Bayesian optimization with the help of the \textit{scikit-optimize} \cite{scikit-optimize} library in Python. Unlike random search or exhaustive methods such as grid search, Bayesian optimization is based on a probabilistic model to efficiently explore the hyperparameter space. We perform 400 iterations to search for the optimal hyperparameters. The selected hyperparameters are detailed in Supplementary Information (SI), Sec. \textcolor{magenta}{IV A} \cite{SI}.

The performance metrics of the classification model on the training and test datasets, using the Random forest algorithm with Bayesian-optimized hyperparameters, are presented in Table~\ref{tab:Classification_metrics}.
\begin{table}[H]
\caption{Performance metrics for classification using the Random forest model with hyperparameters optimized via Bayesian optimization}
\begin{ruledtabular}
\begin{tabular}{ccc}
Metric & Training & Test \\
\hline
Accuracy & 0.989 & 0.932 \\
Recall & 1.000 & 0.986 \\
Precision & 0.988 & 0.941 \\
F1 & 0.994 & 0.963 \\
\end{tabular}
\label{tab:Classification_metrics}
\end{ruledtabular}
\end{table}
\begin{figure}[H]
    \centering
    \includegraphics[width=1\linewidth]{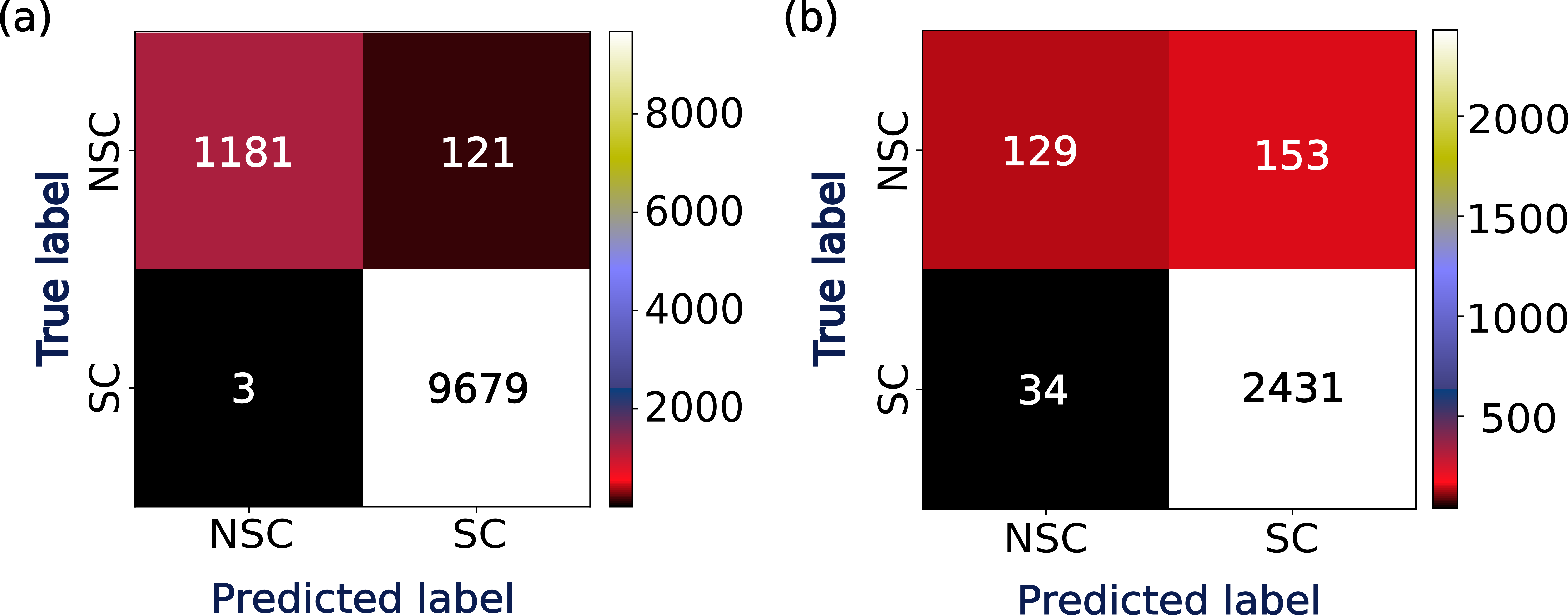}
    \caption{Confusion matrices representing the classification performance for superconductors (labeled as SC) and non-superconductors (labeled as NSC) in the (a) Training and (b) Test datasets.}
    \label{fig:confusion_matrices}
\end{figure}

\vspace{-0.25cm}
\subsubsection{\label{subsubsec: Regression}Regression}
\vspace{-0.25cm}
We select the predicted superconductors using the Random forest classification model, as described in Section~\ref{subsubsec: Classification}. The training and test datasets now consist of 9,800 and 2,584 predicted superconductors, respectively. Bayesian optimization is performed to obtain the best hyperparameters for the regression model used to predict the critical temperature using the Random forest algorithm. The selected hyperparameters are detailed in Supplementary Information (SI), Sec. \textcolor{magenta}{IV B} \cite{SI}.

The performance metrics for the regression model on the training and test datasets, using the Random forest algorithm with Bayesian-optimized hyperparameters, are presented in Table~\ref{tab:Regression_metrics}.
\begin{table}[H]
\caption{Performance metrics for regression using the Random forest model with hyperparameters optimized via Bayesian optimization}
\begin{ruledtabular}
\begin{tabular}{ccc}
Metric & Training & Test \\
\hline
$\text{R}^{2}$ Score & 0.97 & 0.87 \\
MAE (K) & 2.4 & 5.6 \\
RMSE (K) & 4.6 & 10.5 \\
\end{tabular}
\label{tab:Regression_metrics}
\end{ruledtabular}
\end{table}
\begin{figure}[H]
    \centering
    \includegraphics[width=1\linewidth]{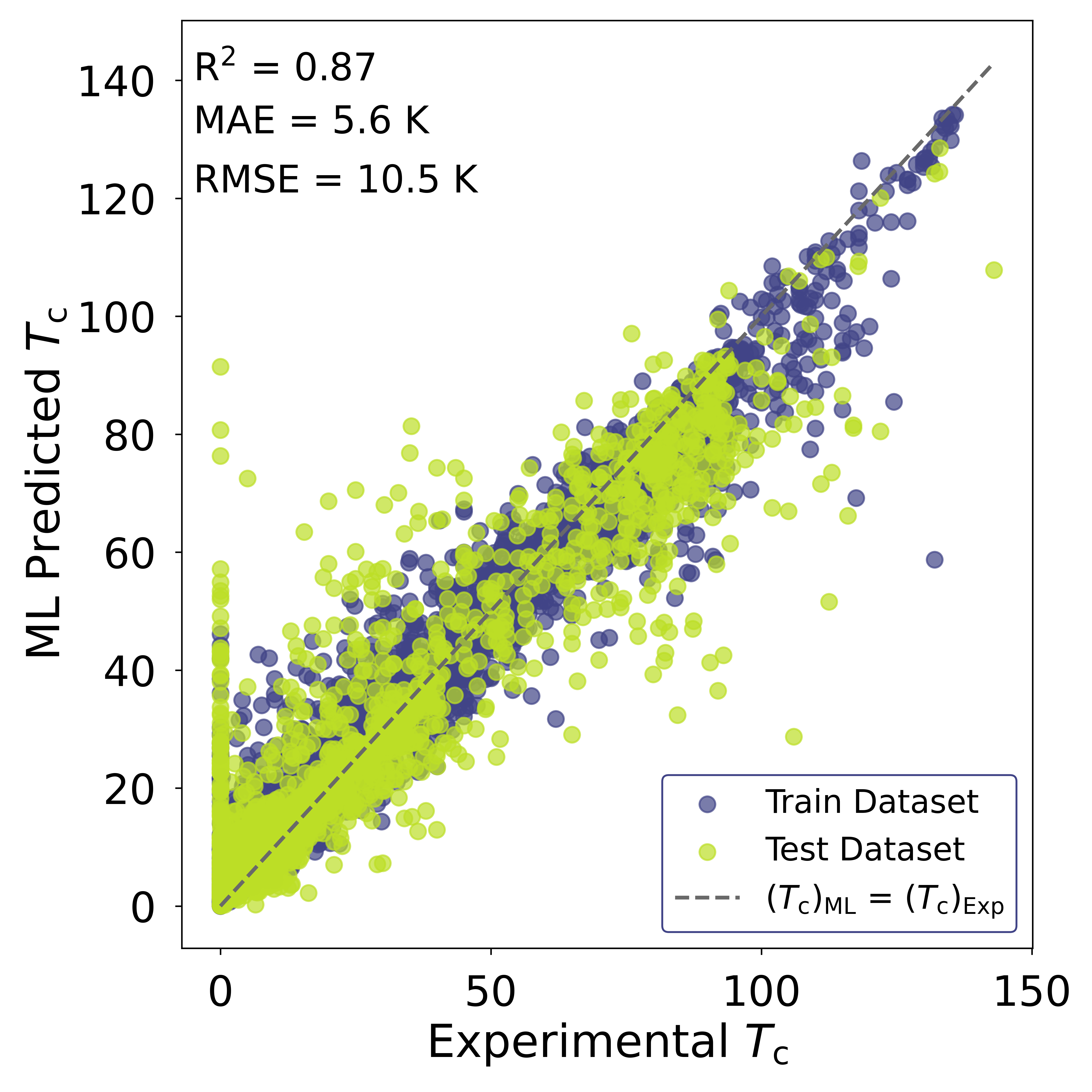}
    \caption{Scatter plot of ML predicted $T_{c}$ versus Experimental $T_{c}$ for compounds classified as superconductors using the Random forest Model.}
    \label{fig:RF_regression}
\end{figure}
It is clear (see Fig.~\ref{fig:RF_regression}) that the ML model learns effectively and performs well on the training dataset. For the test dataset, the model makes accurate predictions but tends to underfit for $T_{c}$ $>$ 40 K. There are abrupt predictions at 0 K, which primarily correspond to non-superconductors misclassified as superconductors by the Random forest classification model. Most of these compounds have compositions near the boundary between superconducting and non-superconducting phases or they belong to the non-superconductors imputed into the dataset reported in \cite{Hosono2015}. We expect these compounds to be superconducting upon tuning of chemical composition or physical parameters like pressure.     

\vspace{-0.25cm}
\subsection{\label{subsec: SHAP Analysis}SHapley Additive exPlanations (SHAP) Analysis of the Random forest Model}
\vspace{-0.25cm}
Random forest is a black-box model, making it difficult to interpret the decision-making process behind its predictions. Explainable AI (XAI) is a field in artificial intelligence that aims to make models more interpretable and explainable. Some of the most commonly used XAI techniques are SHAP \cite{SHAP} and LIME \cite{LIME}.

SHAP explains machine learning model predictions using ideas from game theory. It fairly distributes the contribution of each feature to the final predictions by calculating Shapley values \cite{Shapley1951}, which represent each feature’s average marginal contribution across all possible coalitions. A higher Shapley (SHAP) value indicates that the feature increases the model's prediction relative to the baseline, which is the average of the predicted value over the training dataset. The baseline value for our model in 23.6 K.
\begin{center}
    \includegraphics[width=1\linewidth]{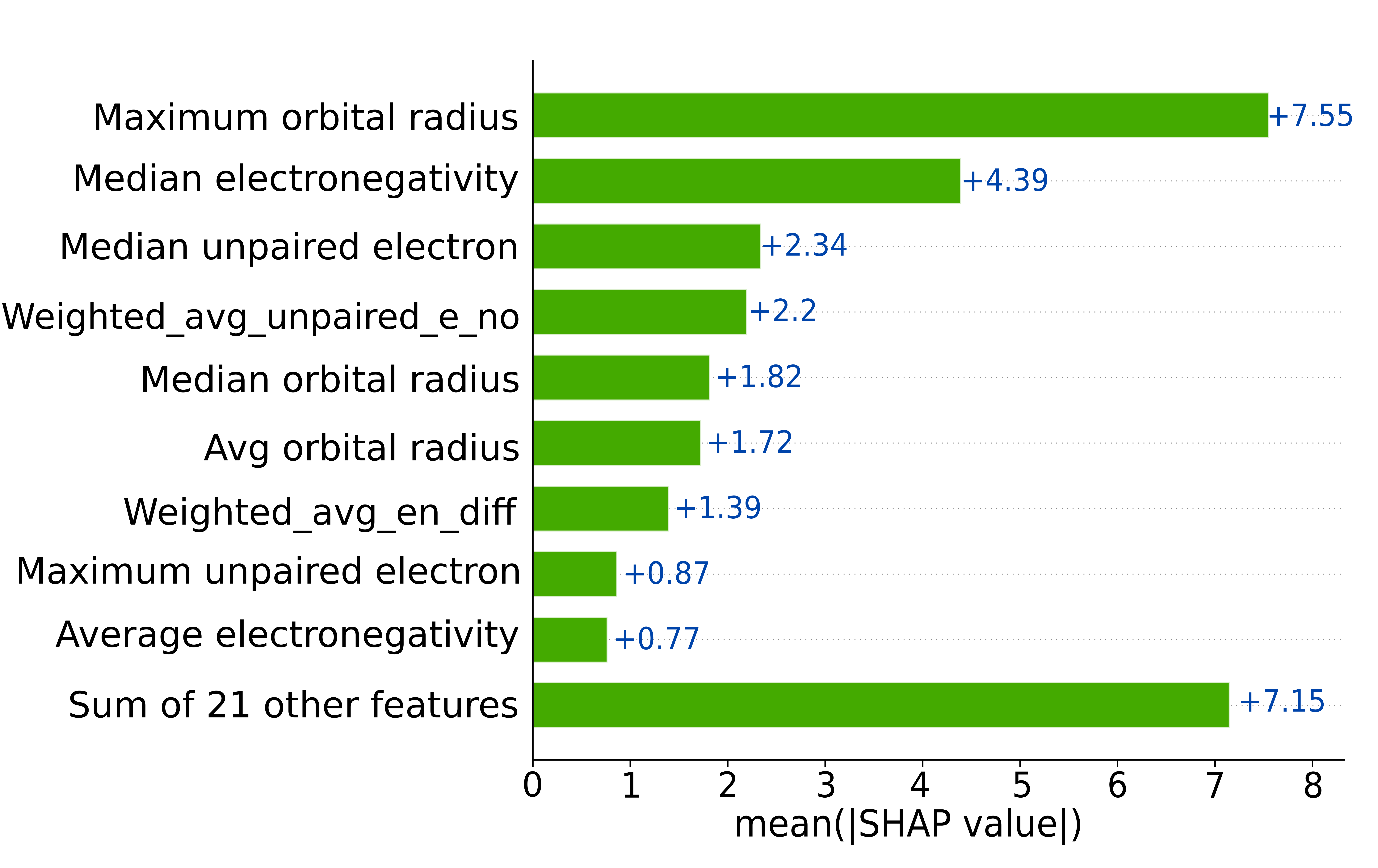}
    \captionof{figure}{Bar plot of the mean absolute SHAP values, with features sorted in decreasing order of their importance.}
    \label{fig:shap_plot}
\end{center}
In Fig.~\ref{fig:shap_plot}, we present a bar plot of the mean absolute SHAP values for each feature, which indicates the ranking of feature importance. A higher mean absolute SHAP value signifies a greater impact on the model’s predictions. However, it does not indicate whether an increase in the feature value leads to a positive or negative SHAP value. `Maximum orbital radius' is the most important feature driving critical temperature prediction, followed by `Median electronegativity', while the last bar represents the contribution of the remaining 21 features.

In Fig.~{\ref{fig:violin_plot}}, we present feature distributions displayed as violin plot illustrating the relationship between feature values and SHAP values. The feature `Maxima Orbital Radius', which has the most significant influence on critical temperature, tends to decrease the critical temperature from the baseline value as the feature value increases. Conversely, `Median Orbital Radius' exhibits a negative SHAP value for higher feature values and a positive SHAP value for lower feature values. This indicates that lower values of this feature increase the predicted critical temperature above the baseline, while higher values decrease it below the baseline.

In the specific case of the "maximum orbital radius" descriptor, this trend aligns with physical insights from copper-oxide based superconductors. A lower orbital radius means a smaller spatial extent of valence orbitals and weaker orbital overlap. This results in the reduced hopping amplitudes and lower bandwidth, and hence stronger electronic correlation. Similarly, for many conventional superconductors, larger orbital radii result in greater overlap and wider, more dispersed bands, reducing the electronic density of states at the Fermi level, which—via the Eliashberg relation—can lead to lower $T_c$. However, this trend could also reflect the fact that more dispersed bands lead to stronger electron–phonon coupling and higher-frequency phonons, both of which have a positive effect on $T_c$. Thus, the feature "maximum orbital radius” has a mixed effect on the conventional class of superconductors. This indicates that the same feature behaves differently across various classes of superconductors. Consequently, the SHAP trend for this feature is difficult to interpret in a physically consistent way across the full dataset.
       
Additionally, average electronegativity has been shown to correlate with $T_c$~\cite{Asokamani1989, Jayaprakash1993}. For example, copper-oxide-based superconductors typically exhibit electronegativity values between 2.12 and 2.53, and our SHAP analysis reveals its positive contributions to $T_c$ within this range. In contrast, elemental superconductors often lie in the 1.3–1.9 range, where SHAP values are predominantly negative, indicating suppression of $T_c$. 

Thus, physically interpretable connections between the statistical descriptors and underlying superconducting mechanisms based on SHAP analysis reinforces the value of combining data-driven approaches to obtain insights for physical understanding.

\vspace{-0.25cm}
\subsection{\label{subsec : 5-Key_Features} Random forest model performance with 5 Key descriptors}
\vspace{-0.25cm}
The top five features influencing the model predictions were picked based on the SHAP analysis, which accounts to 60 \% of model's predictive power, as described in Section~\ref{subsec: SHAP Analysis}, and are:
\begin{enumerate}
    \item Maximum orbital radius
    \item Median electronegativity
    \item Median unpaired electron number 
    \item Weighted average of unpaired electron number 
    \item Median orbital radius
\end{enumerate}
A Random forest regression model was trained for predicting critical temperature with these five key features for materials classified as superconductors by Random forest classification model in Section \ref{subsubsec: Classification}. Hyperparameters for Random forest regression model with 5 Key features are obtained by performing Bayesian Optimization and selected parameters are detailed in Supplementary Information (SI), Sec. \textcolor{magenta}{IV C} \cite{SI}.

The performance metrics for the train and test datasets for critical temperature prediction using the Random forest model with five key features are presented in Table~\ref{tab:Regression_metrics_5KF}. The results demonstrate performance comparable to that of the full model with 30 features, as shown in Table~\ref{tab:Regression_metrics}.
\begin{table}[H]
\caption{Performance metrics for regression using the Random forest model with five key features with hyperparameters optimized via Bayesian optimization}
\begin{ruledtabular}
\begin{tabular}{ccc}
Metric & Training & Test \\
\hline
$R^{2}$ Score & 0.96 & 0.85 \\
MAE (K) & 3.2 & 6.2 \\
RMSE (K) & 5.9 & 11.3 \\
\end{tabular}
\label{tab:Regression_metrics_5KF}
\end{ruledtabular}
\end{table}

\onecolumngrid
\begin{center}
    \includegraphics[width=0.9\linewidth]{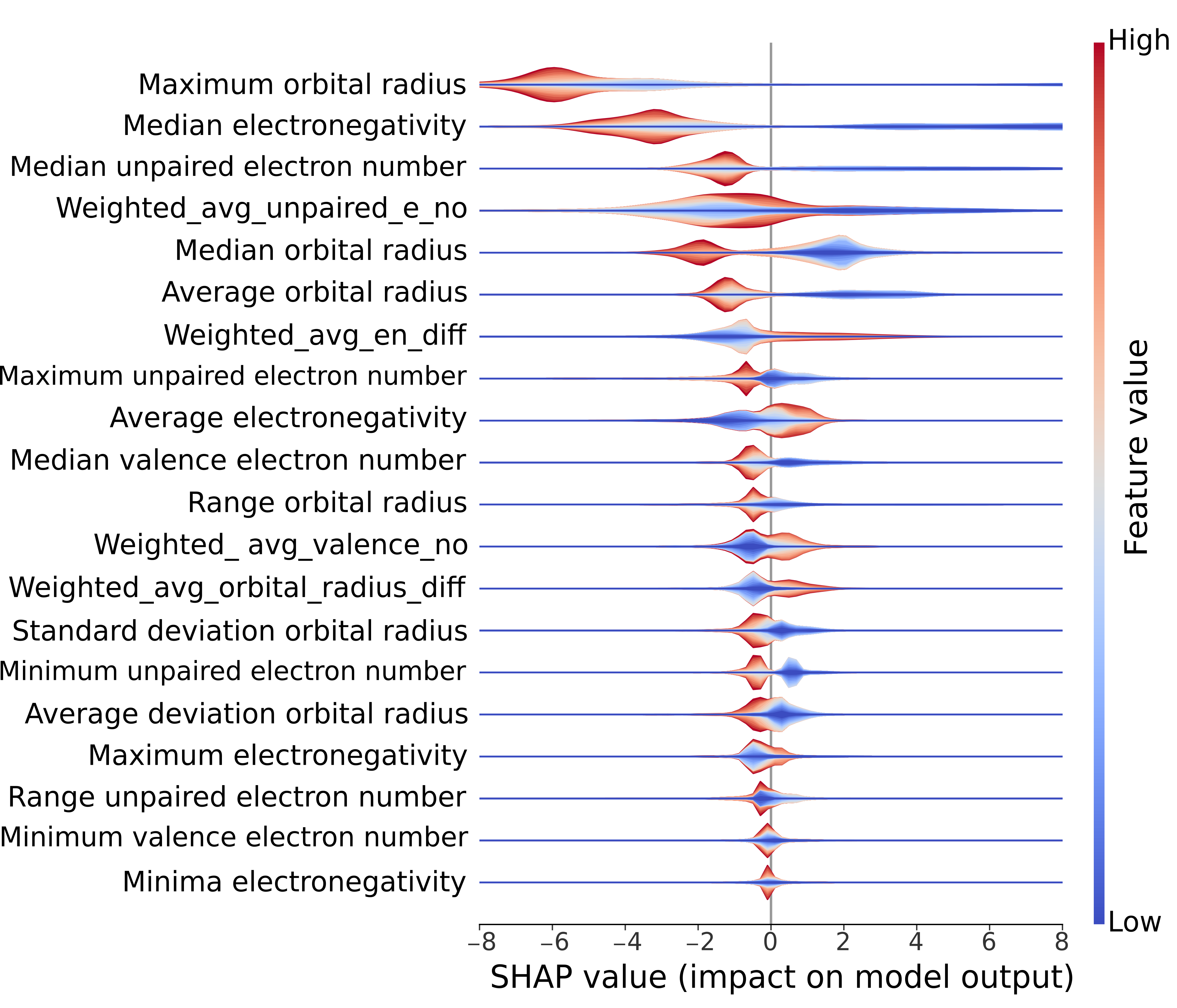}
    \captionof{figure}{Violin plot of feature values and SHAP values.}
    \label{fig:violin_plot}
\end{center}
\twocolumngrid

\onecolumngrid
\begin{center}
    \includegraphics[width=0.88\linewidth]{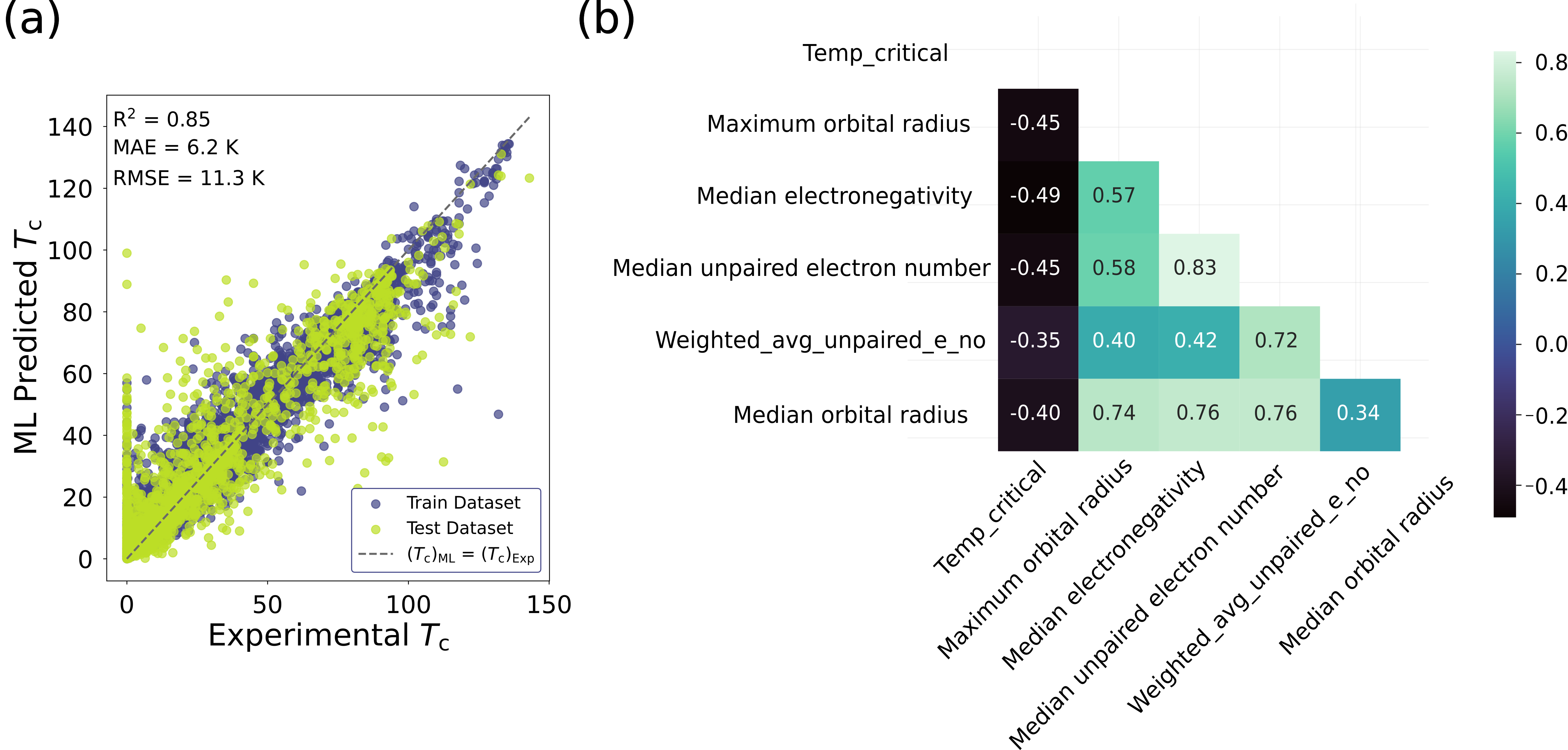}
    \captionof{figure}{(a) Scatter plot of machine learning predicted $T_{c}$ v/s Experimental $T_{c}$ using five key features for compounds classified as superconductors using the Random forest model. (b) Pearson correlation matrix of five key features with critical temperature, showing negative correlation.}
    \label{fig:RF_5_HT}
\end{center}
\twocolumngrid

\vspace{-0.25cm}
\subsection{\label{subsec : Classwise_performance} Classwise performance of Machine Learning Model}
\vspace{-0.25cm}
Around 12,384 compounds are classified as superconductors in \textit{SuperCon-MTG} using Random forest classification model in Sec \ref{subsubsec: Classification} . These materials fall into 24 distinct classes based on the chemical composition. We train a Random forest regression model to predict critical temperature for each class of these compounds, to improve the reliability of ML model predictions, we merge smaller classes with fewer than 170 entries, ensuring adequate dataset size for training. More details on the statistics of classes of compounds can be found in Supplementary Information (SI), Section \textcolor{magenta}{V} \cite{SI}.

We combine $\text{BiS}_2$-based materials and bismuthates into a new class called \textit{Bi-based compounds}. Additionally, A15, Chevrel, Dichalcogenides, Elemental superconductors, Hydrides, Other Iron based materials and unclassified materials are grouped into a single class labeled as the \textit{Rest of the compounds}.  

The $R^2$ score and normalized error, defined as the ratio of the root mean squared error to the average critical temperature of each class, are as shown in Fig.~\textcolor{magenta}{10}. We observe that the Random forest model predicts the critical temperature reasonably well for \textit{Bi-based materials}~\cite{Sleight2015} and \textit{borocarbides}~\cite{Mazumdar2015}, both of which are potentially unconventional superconductors. Similarly, it performs well for \textit{transition metal pnictides}~\cite{Klemm2015} and \textit{borides} \cite{Chen2024}, which are considered conventional superconductors. On the other hand, the accuracy of model predictions for $T_{c}$'s of other classes of superconductors such as \textit{cuprates, nitrides, oxides, and iron-based pnictides and chalcogenides}, are relatively lower.  In case of cuprate superconductors, the optimal Cu–O bond length is around 1.9~Å, and deviations from this value are known to suppress superconductivity. Similarly, the apical Cu–O bond length has been shown to correlate with $T_c$~\cite{KAMBE2000}. In Iron-based materials, structural features such as the \textit{anion height} (the vertical distance between the Fe atom and the anion plane) plays a critical role in superconductivity. An optimal anion height of approximately 1.38~Å has been empirically linked to the highest $T_c$ values in many FeAs-based superconductors~\cite{Chun2014}. For example, in the 1111 phase, replacing La with smaller rare-earth elements such as Nd or Sm increases the anion height from 1.33~Å to 1.38~Å and raises $T_c$ from 26~K to 56~K. However, increasing the anion height beyond this optimal value leads to a decrease in $T_c$, indicating a non-monotonic relationship. This effect is structurally linked to the Fe–pnictogen/chalcogen bond length, which influences both the anion height and the Fe–Pn/Ch–Fe bond angles. The highest $T_c$ values are observed when the FePn$_4$ tetrahedron approaches ideal geometry, with bond angles close to $109.47^\circ$. This structural configuration optimizes orbital hybridization and supports strong spin fluctuation–mediated pairing, which is believed to be responsible for superconductivity in these materials. This indicates that the class-wise trained model performs well for classes where compositional trends dominate, while its performance slightly declines for classes where crystal structure may play a more significant role. Moreover, composition-only features do not capture carrier concentration, magnetic ordering, or phonon properties that do influence superconductivity. Previous machine-learning studies~\cite{Zhangr12020, Zhangr12022} have achieved reasonably realistic $T_{c}$ predictions by incorporating synthesis-specific parameters—such as sintering temperature, time, or pulsed-layer-deposition conditions. Overall, our model shows reasonably good performance and generalizes $T_{c}$ prediction effectively, when trained across classes using only composition-based features. Incorporating structure- and synthesis- specific features among descriptors as variables of predictive models is a promising direction for future work and may significantly improve the model performance, particularly in the high-$T_c$ predictions.
\onecolumngrid
\begin{center}
    \includegraphics[width=1\linewidth]{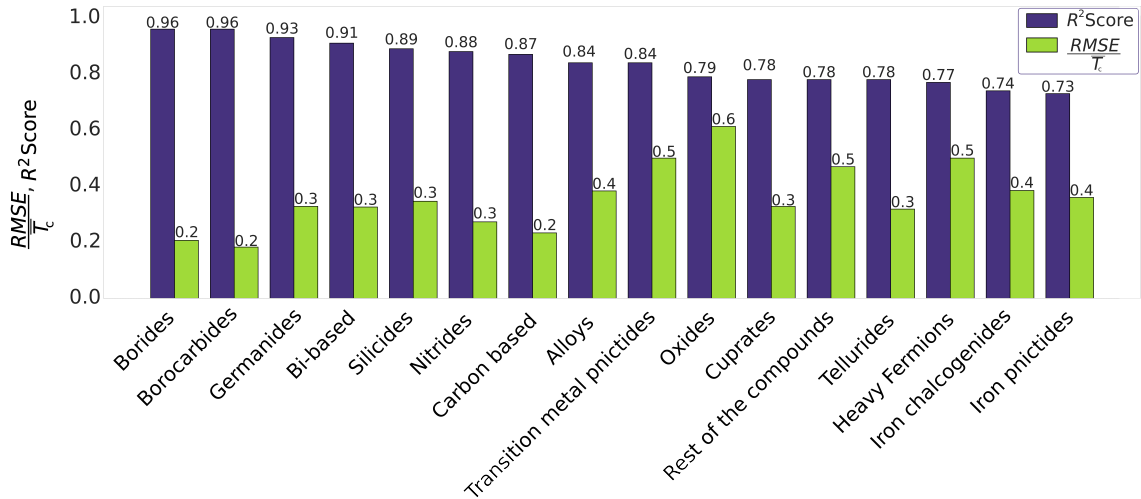}
    \captionof{figure}{Performance metrics for regression using a Random forest model (with default parameters in scikit-learn \cite{scikit-learn}) across different classes of superconducting materials.}
    \label{fig:Model_Performance_class}
\end{center}
\newpage
\twocolumngrid

\vspace{-0.25cm}
\section{\label{sec: New Materials}Validation of our model and Search for new Superconductors}
\vspace{-0.25cm}

\subsection{\label{subsec: SC_Literature} Superconductors from the recent literature}
\vspace{-0.25cm}
In the previous section, we demonstrated that our machine learning model is reasonably robust in classifying materials as superconductors and predicting their $T_{c}$. To further validate its performance beyond the \textit{SuperCon-MTG} dataset, we evaluate materials reported after 2016 that are not included in the training or test datasets. We select 27 compounds that are known to exhibit superconductivity. Our model successfully classifies all of these 27 materials as superconductors and provides reasonably accurate critical temperature predictions. The results are presented in Table~\ref{tab:Superconductors_Literature}. The $\text{Model}_{30}$ and $\text{Model}_{5}$ columns represent $T_c$ predictions using machine learning models with 30 and 5 descriptors, respectively. Similarly, we evaluate our model on 31 compounds previously reported as potential superconductors in earlier machine-learning works~\cite{Jung2024,Pereti2023,Gashmard2024,Kaplan2025}. This benchmarking demonstrates that our model attains performance on par with these prior approaches while employing a smaller, QSD-inspired descriptor set (see Supplementary Information Sec.~\textcolor{magenta}{VIII}~\cite{SI} for details).

\begin{table}[h]
\vspace{0.25cm}
\caption{Machine learning predicted and experimentally reported ($T_c$) of superconducting materials reported in the literature but not included in the \textit{SuperCon-MTG} dataset.}
\centering
\setlength{\tabcolsep}{5pt} 
\begin{adjustbox}{width=1\columnwidth}
\begin{tabular}{lccccl}
\hline\hline
Compound & $\text{Model}_{30}$(K) & $\text{Model}_{5}$(K) & Exp $T_{c}$(K) & Class \\
\hline
$\text{BaB}_{5}$ & 16.3 & 17.6 & 16.3 \cite{Xie2023} & Borides \\ 
$\text{RhSeTe}$ & 4.8 & 4.1 & 4.72 \cite{Patra2024} & Tellurides \\ 
$\text{Re}_{7}\text{Nb}_{3}$ & 5.4 & 6 & 5.2 \cite{Kushwaha2024} & Alloys\\ 
$\text{ZrTe}_{1.8}$ & 2.8 & 3.7 & 3.2 \cite{Correa2023} & Tellurides\\ 
$\text{Re}_{7}\text{Ti}_{3}$ & 4.4 & 5.3 & 3.8 \cite{Kushwaha2024} & Alloys \\ 
$\text{TiIrGe}$ & 3 & 3.4 & 2.24 \cite{Meena2025} & Germanides \\ 
$\text{Na}_{2}\text{CoSe}_{2}\text{O}$ & 7 & 5.5  & 6.3 \cite{Cheng2024} & Oxides  \\
$\text{Re}_{7}\text{Zr}_{3}$ & 4.8 & 4.6 & 5.6 \cite{Kushwaha2024}& Alloys \\ 
$\text{Pt}_{1.8}\text{Ir}_{1.2}\text{Zr}_{5}$ & 4.5 & 3 & 3.2 \cite{Watanabe2023} & Alloys \\ 
$\text{Re}_{7}\text{Hf}_{3}$ & 4.7 & 4 & 6 \cite{Kushwaha2024} & Alloys \\ 
$\text{Re}_{7}\text{Ta}_{3}$ & 5 & 5.5 & 3.6 \cite{Kushwaha2024} & Alloys\\ 
$\text{TaIr}_{2}\text{B}_{2}$ & 3.6 & 2.7 & 5.2 \cite{Gornicka2020} & Borides \\ 
$\text{Pt}_{1.5}\text{Ir}_{1.5}\text{Zr}_{5}$ & 4.9 & 4.9 & 3.2 \cite{Watanabe2023} & Alloys \\ 
$\text{Pt}_{1.2}\text{Ir}_{1.8}\text{Zr}_{5}$ & 4.5 & 5.1 & 2.8 \cite{Watanabe2023} & Alloys\\
$\text{Cu}_{2}\text{Si}$ & 2.1 & 2.7 & 4.1 \cite{Yan2019} & Silicides \\ 
$\text{HfRhGe}$ & 3.8 & 4 & 1.7 \cite{P2024} & Germanides\\
$\text{Pt}_{2.7}\text{Ir}_{0.3}\text{Zr}_{5}$ & 6 & 2.5 & 3.8 \cite{Watanabe2023} & Alloys \\ 
$\text{Mo}_{20}\text{Re}_{6}\text{Si}_{4}$ & 7.7 & 12.3 & 5.4 \cite{Kaplan2025} & Silicides\\ 
$\text{Pt}_{2.1}\text{Ir}_{0.9}\text{Zr}_{5}$ & 5.8 & 3.5 & 3.4 \cite{Watanabe2023} & Alloys\\ 
$\text{Pt}_{0.9}\text{Ir}_{2.1}\text{Zr}_{5}$ & 5.1 & 5.5 & 2.7 \cite{Watanabe2023} & Alloys\\ 
$\text{Pt}_{2.4}\text{Ir}_{0.6}\text{Zr}_{5}$ & 6.2 & 3.9 & 3.6 \cite{Watanabe2023} & Alloys\\ 
$\text{HfIrGe}$ & 3.1 & 3.5 & 5.64 \cite{Meena2025}  & Germanides\\ 
$\text{Pt}_{0.45}\text{Ir}_{2.55}\text{Zr}_{5}$ & 5.3 & 5.9 & 2.6 \cite{Watanabe2023} & Alloys \\ 
$\text{Pt}_{0.6}\text{Ir}_{2.4}\text{Zr}_{5}$ & 5.3 & 5.9 & 2.6 \cite{Watanabe2023} & Alloys\\ 
$\text{NbIr}_{2}\text{B}_{2}$ & 4.2 & 3.2 & 7.2 \cite{Gornicka2020} & Borides\\ 
$\text{Pt}_{3}\text{Zr}_{5}$ & 3.5 & 2.8 & 7.3 \cite{Hamamoto2018} & Alloys\\ 
$\text{Re}_{8}\text{NbTa}$ & 3.4 & 3.9 & 7.7\cite{Kushwaha2024-1} & Alloys\\ 
$\text{ReTc}$ & 6.9 & 6.3 & 11.5 \cite{Cerqueira2023}\footnote{DFT predicted \( T_c \)} & Alloys \\
\end{tabular}%
\label{tab:Superconductors_Literature}
\end{adjustbox}
\end{table}

\vspace{-0.25cm}
\subsection{\label{subsec: Materials_Project} Using the compounds in the Materials Project}
\begin{figure}[H]
    \centering
    \includegraphics[width=1\linewidth]{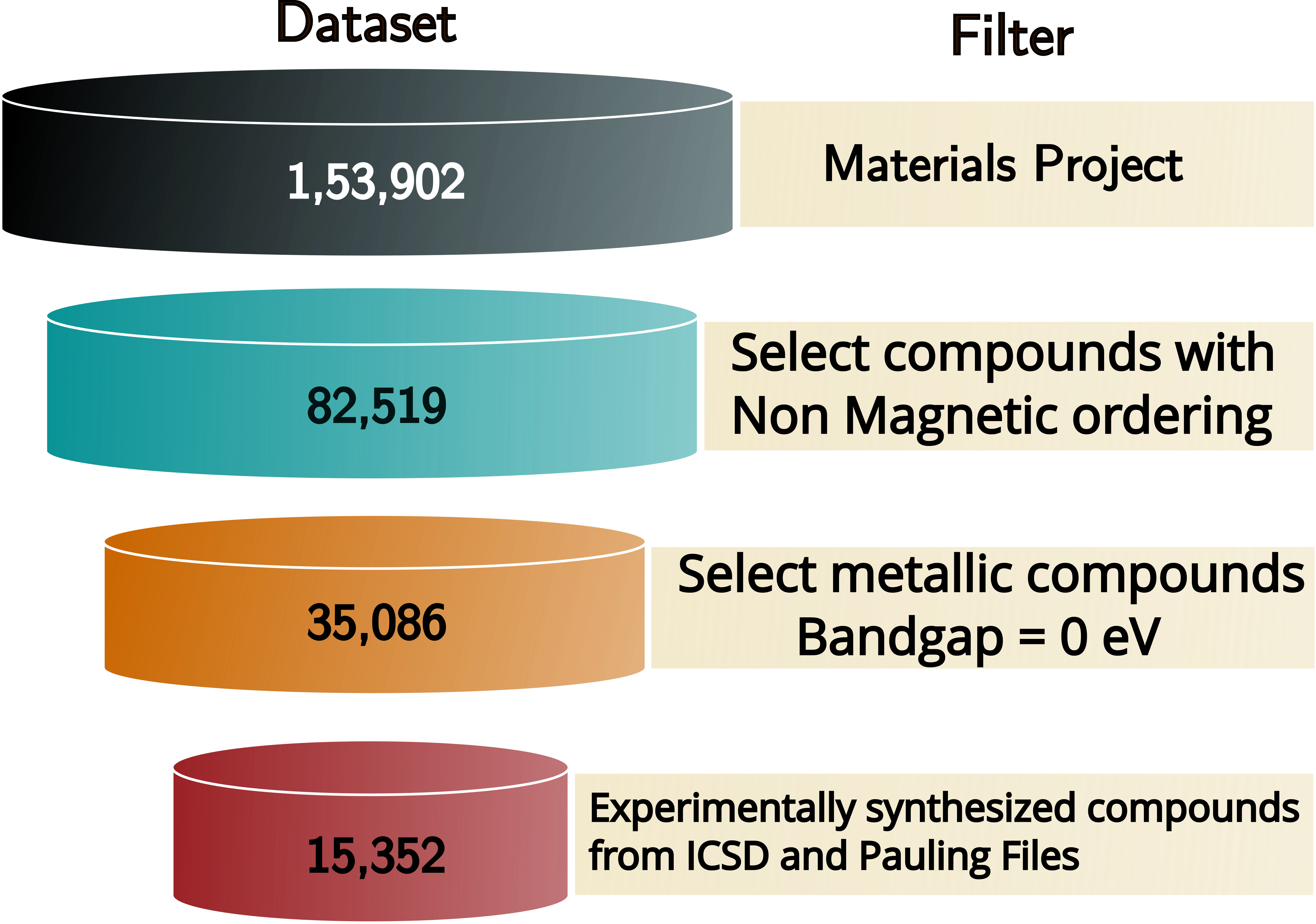}
    \caption{A schematic of the filtering process applied to the Materials Project database in search of potential superconductors using machine learning models.}
    \label{fig:MP_filter}
\end{figure}

\vspace{-0.25cm}
To identify potential superconducting materials, we screen compounds from the Materials Project database \cite{Jain2013}, which hosts over 150,000 inorganic materials. We refine our selection by filtering for nonmagnetic, metallic, and experimentally reported compounds that are also present in the ICSD \cite{Belsky2002} and Pauling File database \cite{Villars2019}, as illustrated in Fig.~\ref{fig:MP_filter}. This reduces the dataset to approximately 15,352 compounds, of which around 1,606 are already listed in \textit{SuperCon-MTG}, leaving approximately 13,746 candidate materials. We apply our machine learning model to this refined dataset to identify potential superconducting materials. The top 10 compounds with the highest predicted critical temperatures, as determined using both the 30-feature and 5-feature models, are presented in Table~\ref{tab:Superconductors_MP}. 4 of these 10 compounds turn out to be reported superconductors, but not present in \textit{SuperCon-MTG}.  The predicted critical temperatures for $\text{Tl}_2\text{Ba}_2\text{Ca}_3\text{Cu}_4\text{O}_{12}$, $\text{Er}\text{Ba}_2\text{Cu}_4\text{O}_8$, and $\text{Tl}_2\text{Sr}_2\text{Ca}\text{Cu}_2\text{O}_{7}$ from both the 30-feature and 5-feature models are very close to the corresponding experimental values. For $\text{Bi}_2\text{Sr}_2\text{Ca}_3\text{Cu}_4\text{O}_{12}$, the prediction from the 30-feature model aligns well with the experimental $T_c$, while the 5-feature model underestimates it, though it still successfully identifies the compound as a superconductor. The remaining six materials are new predictions and should be tested for superconductivity. Most of the compounds listed in Table~\ref{tab:Superconductors_MP} contain copper in the +2 oxidation state. In $\text{Tl}_5\text{Ba}_6\text{Ca}_6\text{Cu}_9\text{O}_{29}$, copper exists in both +1 and +2 oxidation states. In $\text{Er}\text{Ba}_2\text{Cu}_4\text{O}_8$, the average oxidation state of copper is +2.25, with three copper atoms in the +2 state and one in the +3 state. The presence of copper in the +2 state suggests that hole doping is required for the material to exhibit superconductivity.

One limitation of our model is that it is mostly trained on materials which are metallic or possess a small band gap. So, our predictions are expected to be most reliable for compounds with metallic or narrow-gap materials.

\onecolumngrid
\begin{center}
\begin{table}[H]
\caption{Top 10 potential superconducting materials identified from the Materials Project database, sorted in descending order of predicted critical temperature ($T_c$).}
\begin{ruledtabular}
\begin{tabular}{cccccc}
Material-ID & Compound & $\text{Model}_{30}$(K) & $\text{Model}_{5}$(K) & Experimental $T_{c}$(K) & Class \\
\hline
mp-680433 & $\text{Tl}_5\text{Ba}_6\text{Ca}_6\text{Cu}_9\text{O}_{29}$ & 104.6 & 104.2 & -  & Cuprates\\ 
mp-556574 & $\text{Tl}_2\text{Ba}_2\text{Ca}_3\text{Cu}_4\text{O}_{12}$ & 101.6 & 98.5 & 108 \cite{Dingyun1989} & Cuprates \\
mp-542197 & $\text{Tl}_3\text{Ba}_4\text{Ca}_4\text{Cu}_6\text{O}_{19}$ & 98.7 & 93.5 & - & Cuprates \\ 
mp-1191679 & $\text{Bi}_2\text{Sr}_2\text{Ca}_3\text{Cu}_4\text{O}_{12}$ & 82.1 & 43.9 & 90 \cite{Matheis1990} & Cuprates\\ 
mp-1208803 & $\text{HgLa}\text{Sr}_2\text{Cu}_2\text{O}_6$ & 75.8 & 67 & - & Cuprates\\ 
mp-6583 & $\text{Er}\text{Ba}_2\text{Cu}_4\text{O}_8$ & 74.5 & 71.5 & 78.5 \cite{Fischer1992} & Cuprates\\ 
mp-1208800 & $\text{Bi}_4\text{Sr}_2\text{Ca}_4\text{Cu}_4\text{O}_{16}$ & 74 & 75.4 & - & Cuprates\\ 
mp-1209015 & $\text{Bi}_2\text{Sr}_2\text{Ca}_2\text{Cu}_3\text{O}_{10}$ & 73.5 & 75.3 & - & Cuprates\\ 
mp-1208765 & $\text{Tm}_2\text{Ga}_2\text{Sr}_4\text{Cu}_4\text{O}_{14}$ & 68 & 78.5 & - & Cuprates\\ 
mp-20824 & $\text{Tl}\text{Sr}_2\text{Ca}\text{Cu}_2\text{O}_{7}$ & 67 & 69 & 80 \cite{AbdShukor1996} & Cuprates\\ 
\end{tabular}
\label{tab:Superconductors_MP}
\end{ruledtabular}
\end{table}
\end{center}

\twocolumngrid
\vspace{-0.25cm}
\subsection{\label{subsec: Unexplored_space} Unexplored Space of materials}
\vspace{-0.25cm}
Manual search for potential superconducting materials within the Materials Project database is rather a time intensive and impractical process. To streamline this process, we use the SMACT library \cite{Davies2019} in Python to generate binary compositions consisting of one transition metal and one main group element (details provided in Supplementary Information, Sec. \textcolor{magenta}{VI} \cite{SI}). Elements from the 7th period, noble gases, and radioactive elements such as Tc, Po, and At are excluded. The generated compositions are then validated based on fundamental chemical principles, ensuring charge neutrality and conditions on electronegativity.

Furthermore, we compute the probability of these generated compounds to exist using the given oxidation states of constituents (see Supplementary Information Sec \textcolor{magenta}{VII} \cite{SI}) and retain only those with a probability greater than zero. This filtering process helps to retain chemically plausible compositions and reduces the initial 5,085 possible compositions to 292, of which 19 are already listed in \textit{SuperCon-MTG}. We further refine the dataset of remaining 273 compounds by applying the filtering criteria outlined in Fig.~\ref{fig:MP_filter}. From this, we shortlist 11 materials, 4 of which have already been reported as superconductors, predominantly in studies published after 2016, as presented in Table~\ref{tab:Superconductors_Binary}. In addition, we propose 7 new compounds as potential superconducting materials, as shown in Table~\ref{tab:Potential_Superconductors_Binary}.
\onecolumngrid
\begin{center}
\begin{table}[H]
\caption{Experimental confirmation of machine learning model predictions for materials with compositions generated using the SMACT library.}
\begin{ruledtabular}
\begin{tabular}{cccccc}
ICSD & Compound & $\text{Model}_{30}$(K) & $\text{Model}_{5}$(K) & Experimental $T_{c}$(K) & Class \\
\hline
657372 & $\text{Ta}_{2}\text{Se}$ & 4.2  & 8.1  & 3.8 \cite{Gui2020} & Others  \\
646899 & $\text{NiTe}$ & 2.7 & 2.6 & 1.5 \cite{Liang2023}\textcolor{magenta}{\textsuperscript{b}} &  Tellurides \\
238590 & $\text{FeS}$ & 5.8 & 7 & 5 \cite{Lai2015} & Iron Chalcogenides \\
- & $\text{NbS}$ & 4.4 & 6.7 & 5.6 \cite{Ruan2023} & Others  \\
\end{tabular}
\label{tab:Superconductors_Binary}
\end{ruledtabular}
\end{table}
\hspace{-14cm}
\noindent\textsuperscript{b}DFT predicted \( T_c \).
\begin{table}[H]
\caption{Compositions generated using SMACT and predicted to be superconducting by the machine learning model, along with their predicted $T_c$.}
\begin{ruledtabular}
\begin{tabular}{ccccc}
ICSD & Compound & $\text{Model}_{30}$(K) & $\text{Model}_{5}$(K) & Class \\
\hline
86115 & $\text{CoTe}_{2}$ & 1.8  & 1.9  &  Dichalcogenides \\
101820 & $\text{Ag}_{2}\text{Te}$ & 1.6  & 1.8  & Tellurides \\
659260 & $\text{IrS}_{2}$ & 4  & 5.6  & Dichalcogenides\\
652055 & $\text{TiSe}$ & 2.9 & 2.9 & Others  \\
56216 & $\text{HgTe}$ & 4.1 & 4.7 & Tellurides \\
41047 & $\text{Ta}_{2}\text{Te}_{3}$ & 3.7 & 4 & Tellurides  \\
42982 & $\text{Ta}_{2}\text{Se}_{3}$ & 4.1 & 4.5 & Others  \\
\end{tabular}
\label{tab:Potential_Superconductors_Binary}
\end{ruledtabular}
\end{table}
\end{center}
\newpage
\twocolumngrid
\vspace{-0.25cm}
\section{\label{sec: Summary} Summary}
\vspace{-0.25cm}
In this work, we present a workflow to modify and refine existing database of superconducting compounds and present a cleaner and curated version. Using ideas from Quantum Structure Diagrams, we demonstrate clustering of superconducting compounds by their classes in the plane of principal components revealing meaningful structure in the data. Building on this foundation, we developed a machine learning framework for efficient classification of materials as superconductor, and prediction of its $T_{c}$ directly from its chemical composition. Remarkably, the model requires no crystallographic information and relies only on Quantum Structure Diagram based descriptors and Random forest classification and regression algorithms. Importantly, the model generalizes well beyond the \textit{SuperCon-MTG} dataset, accurately predicting superconductivity in newly discovered compounds for superconductivity. We also identify several promising candidates for superconductivity, which will stimulate new experiments. This showcases our model's potential to accelerate the discovery of novel superconducting materials.

However, our model is limited by the lack of reliable non-superconductor data, which introduces dataset bias and constraints the development of more accurate $T_c$ prediction models. The use of composition-only descriptors further restricts the framework by omitting key physical factors underlying superconductivity. Incorporating crystal-structure information and synthesis conditions available from the literature, represents a natural next step toward building more accurate machine-learning models and advancing our understanding of superconductivity through materials informatics.

\vspace{-0.25cm}
\section{Acknowledgments}
\vspace{-0.25cm}
S.A. acknowledges JNCASR for a research fellowship. The authors thank Prof. Ram Seshadri for constructive suggestions and comments on manuscript. We thank Prof. A. Sundaresan and Prof. Ricardo Grau-Crespo for valuable discussions and insights. U.V.W. acknowledges support from J. C. Bose Grant (2025/0210) of ANRF, Government of India.

\vspace{-0.25cm}
\section{Code and Data availability}
\vspace{-0.25cm}
The \textit{SuperCon-MTG} dataset, trained models, and code for feature generation and critical temperature prediction are available in the \href{https://github.com/adigasuhas/Accelarating-Search-for-Superconductors-using-Machine-Learning.git}{GitHub repository}: \url{https://github.com/adigasuhas/Accelarating-Search-for-Superconductors-using-Machine-Learning.git}

\bibliographystyle{apsrev4-2} 
\bibliography{References}
\end{document}